\documentclass[journal]{IEEEtran}

\usepackage{times,amsmath,amssymb,epsfig}

\begin{document}

\title{An Integral-Equation Method Using Interstitial Currents Devoted
to the Analysis of Multilayered Periodic Structures 
with Complex Inclusions}

\author{Nilufer~A.~Ozdemir,~\IEEEmembership{Member,~IEEE,}
        and~Christophe~Craeye,~\IEEEmembership{Senior Member,~IEEE}}      
\thanks{N.~A.~Ozdemir and C~Craeye are with the ICTEAM, Universit\'e Catholique de Louvain,
Louvain-la-Neuve, B-1348 Belgium e-mail: \{nilufer.ozdemir,christophe.craeye\}@uclouvain.be.}
\thanks{Manuscript received August 10, 2014.}

\markboth{Submitted to IEEE Transactions on Antennas and Propagation, August 9, 2014}%
{Shell \MakeLowercase{\textit{et al.}}: Bare Demo of IEEEtran.cls for Journals}

\maketitle

\begin{abstract}
An efficient surface integral equation-based method is proposed for the analysis of electromagnetic scattering 
from multilayered media containing complex periodic inclusions. The proposed method defines equivalent currents
at the interfaces between layers in order to eliminate the need to compute the layered medium Green's function. 
Hence, the background medium in a given layer can be treated as a homogeneous unbounded medium for which 
the computation of the Green's function for an infinite doubly periodic array is sufficient. 
The resulting method-of-moments interaction matrix has a block tridiagonal structure,
which leads to computational complexity proportional to the number of layers for both matrix filling and solution.
When all layers are identical, the filling time essentially reduces to that of a single layer, and the
interaction matrix has a Toeplitz structure. Numerical results are provided for the reflectivity of multilayered 
periodic arrays of spherical silver core-silica shell nanoparticles, excited by a plane wave at optical frequencies. 
Comparisons with results obtained with an FDTD-based commercial software validate the accuracy and efficiency 
of the proposed method.
\end{abstract}


%
\IEEEpeerreviewmaketitle

\section{Introduction}
\IEEEPARstart{T}he {electromagnetic} (EM) modeling and analysis of arbitrarily-shaped penetrable or non-penetrable
objects embedded in layered media constitutes an active and important area of research. Such structures find applications 
at both microwave and optical frequencies such as microstrip antennas, microwave circuits, solar cells, lithography, and 
geophysical exploration. When the inclusions form a 2-D periodic lattice in each layer, 
such structures may be used as metamaterials, i.e. materials that may have, 
within certain frequency ranges, properties that cannot be found in nature \cite{smith2000}--\cite{engh2006}.
{Surface integral equation-based methods are advantageous 
for the numerical analysis of electromagnetic scattering from penetrable structures
because, in those methods, the radiation condition is satisfied 
implicitly, and unknowns are limited to interfaces between piecewise homogeneous media. Among possible
surface integral-equation formulations, the PMCHWT approach \cite{journal21}--\cite{journal23},
which explicitly imposes the continuity of tangential electric and magnetic fields across interfaces, is widely employed \cite{rao1990}.
The method-of-moments (MoM) \cite{book1} solution of the PMCHWT formulation has been successfully applied to different periodic 
and non-periodic penetrable materials in a wide range of applications {\cite{usner2007}--\cite{journal40}.}

Traditionally, the integral equation-based numerical solution of this type of problems requires the computation
of the layered medium Green's function in order to take into account the EM effect of layered media. However, obtaining the 
spatial-domain periodic Green's function for a layered medium from its analytically expressed spectral-domain counterpart 
requires the evaluation of Sommerfeld integrals or summations; the former for isolated sources and the latter
for periodic sources. Such evaluations are computationally expensive processes due to the oscillatory and 
slowly decaying nature of the integrand or series \cite{journal1}--\cite{volski2013}. 
{The analysis of multilayered periodic structures using such an approach
 has two additional major drawbacks. First, the computational complexity of matrix filling is proportional to the square of the number 
of layers in the structure. Second, a new analysis is necessary every time a change is made in any layer. 

The Generalized Scattering Matrix (GSM) \cite{journal30} technique provides a solution to the latter two drawbacks.
It describes the reflection and transmission properties of each layer by a scattering matrix, and it uses a cascading process 
to obtain a scattering matrix for the overall structure. To this end, every possible Floquet harmonic
of the periodic incident field is considered, and the transmission and reflection coefficients for all Floquet harmonics are computed.
The scattering matrix of the layered structure can then be obtained from the scattering matrices
of the individual layers. In practice, the maximum order of Floquet harmonics depends on
the period of the structures and on the thickness of the layers.
Different numerical methods, such as the MoM \cite{journal31} and the finite-difference time-domain (FDTD) method \cite{journal32}, 
have been employed to calculate the elements of the scattering matrix.}
 
\indent{The present paper proposes an efficient surface integral equation method for the analysis of multilayered periodic objects
which are either embedded in or located above layered media. The period will be assumed to be the same in each layer,
but the host medium in different layers, their thicknesses and the inclusions they contain can be different.
The proposed method, briefly introduced in \cite{conf11}, is based on surface equivalence used at two different levels. 
First, objects are analyzed via equivalent currents placed on the interfaces between piecewise homogeneous media. 
Second, when the objects are embedded in multiple layers, an equivalence plane, with equivalent electric and magnetic currents, 
is inserted at every interface between layers. Hence, the medium in a given layer can be treated as a homogeneous unbounded 
medium, and the computation and tabulation of the doubly periodic Green's function is sufficient. 
Moreover, the global interaction matrix is Toeplitz when the background medium and the object remain identical in each layer. 
In the MoM solution, the interaction matrix for each layer is independent, and the global interaction matrix has a block tridiagonal form,
inherited from the use of surface equivalence, which also allows the elimination of variables related to the complex inclusions.
Here, contrary to the treatment of non-periodic complex objects \cite{vdwater2005}--\cite{lanc2009}, the exploitation 
of surface equivalence in layered periodic structures allows the reduction of equivalence surfaces to open surfaces with very small area, 
corresponding to the portion of interface between layers that is located within the unit cell. This leads to a very modest
total number of unknowns in the final system of equations. 
The main advantage of the proposed method over the GSM method is that the whole formulation is based
on a minimal set of types of quantities to be determined and of numerical tools to be used. 
Indeed, besides a traditional MoM solver for isolated penetrable bodies \cite{rao1990},
the only building block needed is a code able to test electric and magnetic fields radiated 
by a 2-D periodic array of sources, decomposed into elementary basis functions \cite{usner2007}--\cite{journal37}.
From there, a reduced system of equations, involving  only unknowns related to
equivalent currents at the interfaces, is readily established. 
Those interfaces do not necessarily need to be planar; also, in case the inclusions are nearly touching the interfaces, 
it is in principle possible to refine the mesh on the interfaces in regions very close to the inclusions.}

\indent{The remainder of this paper is organized as follows. 
Section II describes the details of the proposed numerical approach. 
Section III provides numerical results for the reflectivity of infinite 
layers made of doubly periodic arrays of spherical silver core-silica shell 
nanoparticles located above a silicon substrate. This structure corresponds
to an interesting metamaterial with a minimum reflectivity for a wavelength 
near 400 nm \cite{journal34}. The results are validated by comparison with numerical results obtained 
using the commercial software Lumerical \cite{lum} for arrangements comprising up to seven layers. 
Section IV outlines the main contributions and provides further perspectives for this approach.}

\section{Formulation}
\label{sec:form}

The multi-layer strucutre is represented schematically in Fig.~\ref{fig:layers}. There are $N$ {\it interfaces}, 
numbered from $0$ to $N-1$ bottom-to-top, and $N-1$ {\it layers}, numbered from 1 to $N-1$, between the interfaces.
Layer $n$ is just beneath interface $n$, while homogeneous semi-infinite media
above and below the layered structure are considered as layers 0 and $N$. Within the unit cell
of the 2-D periodic structure, each layer 1 to $N$ contains an {\it inclusion} that can consist of several 
disconnected objects. It is assumed that, in each layer, the objects are composed of piecewise
homogeneous materials, and that the host medium is homogeneous.
However, the host media composing
different layers, the thicknesses of the layers, as well as the inclusions in successive layers, can be different. 
It is interesting to notice that, for the formulation given hereafter to be applicable, the interfaces do not need to be flat. 

\begin{figure}[htb]
\begin{center}
\hspace*{-0.75cm}\includegraphics[scale=0.28]{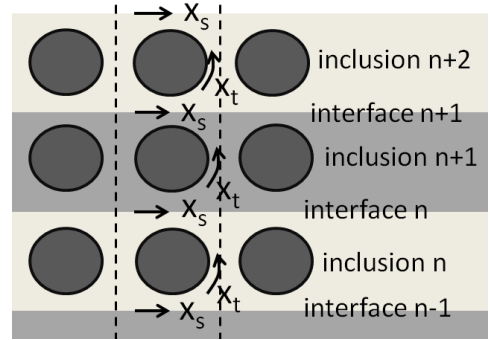}
\caption{Layered structure with inclusions, with numbering of layers and planar interfaces.}
\label{fig:layers}
\end{center}
\end{figure}

The formulation given hereafter relies on the surface equivalence principle. More precisely, 
the PMCHWT formulation \cite{journal21}--\cite{journal23}, \cite{rao1990},
sometimes also named ``continuity formulation'' \cite{handbook}, will be employed. In a nutshell, the PMCHWT approach
defines equivalent electric and magnetic currents on each interface between homogeneous media.
For simplicity, the formulation hereafter will be given for a homogeneous penetrable inclusion;
the appendix briefly describes the extension to more complex inclusions.
The equivalent electric and magnetic currents are $\vec{J} = \hat{n}\times \vec{H}$ and
$\vec{M} = -\hat{n}\times \vec{E}$,  respectively; $\hat{n}$ is the unit normal pointing outward the penetrable 
object, and  $\vec{H}$ and $\vec{E}$ are the magnetic and electric fields. The
equivalent currents are determined by explicitly ensuring the continuity of tangential electromagnetic fields 
across the surface of the object. That continuity also appears implicitly through the definition of opposite equivalent 
currents on both sides of the interface. As a convention, we will assume that the actual unknown currents are those just 
above the interfaces and just outside the inclusions (see Fig.~{\ref{fig:layers}). The continuity of tangential 
$\vec{E}$ and $\vec{H}$ fields will be imposed across both the interfaces and the
boundaries of the inclusions. As will be explained hereafter, only the unknows related to equivalent currents 
defined along the interfaces will appear in the final system of equations, which will be sparse.

The proposed formulation may be regarded as an extension of that described in \cite{journal19}, 
where the case of a single layer containing metallic inclusions is considered. As done in that paper, 
in general, the Green's function used to link sources and magnetic vector potential is the 2-D periodic 
scalar Green's function associated with the host material taken as an unbounded host medium. 
One exception to this rule must be considered: inside the inclusions, the link between equivalent 
currents and radiated fields is obtained using the scalar Green's function for an isolated source
since the interior problem is not periodic. An important aspect of the proposed formulation is that the unknowns 
corresponding to the equivalent currents on the surface of the inclusions, which may sometimes have complex
geometries, will be eliminated from the system of equations.

In order to alleviate the notation, the unknowns associated with the equivalent electric and magnetic currents defined 
on a given interface are represented with a single vector $\bf x$:  
\begin{equation}
{\bf x} = \left[
\begin{array}{cc}
{\bf j}\\
{\bf m}
\end{array}
\right]
\end{equation}
where $\bf j$ and $\bf m$ are column vectors that respectively list the coefficients of equivalent electric and magnetic currents,
with respect to the basis functions. Here, the same set of basis functions will be used to represent electric and magnetic currents. 
Besides, Garlekin testing will be used, in that the set of basis functions also corresponds to the set of testing functions.

The electric and magnetic fields tested by a set of testing functions defined on surface $S_1$ 
and radiated by equivalent currents, described on surface $S_2$ by a set of basis functions with 
coefficients ${\bf x}_2$, are obtained as ${\bf y}_1 = {\bf Z}\,{\bf x}_2$, with
\begin{equation}
{\bf Z} = 
\left[ \begin{array}{cc}
\frac{\eta {\bf Z}^{EJ}}{j\,k} & {\bf Z}^{EM}\\
- {\bf Z}^{EM} & \frac{{\bf Z}^{EJ}}{j\,k\,\eta}
\end{array} \right]
\label{eq:MatrixBlock}
\end{equation}
where $\eta$ and $k$ are the impedance and wavenumber of the medium through 
which basis and testing functions interact, respectively. The $Z^{EJ}$ 
and $Z^{EM}$ matrix blocks are defined by:
\begin{multline}
Z_{k,l}^{EJ}=\int_{S_{1}}\, \int_{S_{2}}\,G(\vec{r}_1,\vec{r}_2)\,
(k^2\vec{F}_{k}(\vec{r}_2) \cdot \vec{F}_{l}(\vec{r}_1)-\\
\nabla \cdot \vec{F}_{k}(\vec{r}_2)
\nabla \cdot \vec{F}_{l}(\vec{r}_1))\, dS_2\, dS_1
\end{multline}
\begin{multline}
Z_{k,l}^{EM}=\int_{S_{1}}\int_{S_{2}}\nabla_1 G(\vec{r}_1,\vec{r}_2)\times
\vec{F}_{k}(\vec{r}_1) \cdot \vec{F}_{l}(\vec{r}_2)\, dS_1\, dS_2
\end{multline}
where $\vec{F}_{l}$ and $\vec{F}_{k}$ are the elementary basis function $l$ and the testing function $k$ on source 
domain $S_2$ and observation domain $S_1$, respectively. $G(\vec{r}_1,\vec{r}_2)$ 
is the scalar Green's function, which depends on the source and observation points $\vec{r}_2$ and $\vec{r}_1$
and on the properties of the material through which basis and testing functions are interacting.
$\nabla_1 G$ is the gradient of $G$ with respect to the $\vec{r}_1$ coordinates.
Function $G$ corresponds to the periodic Green's function in the medium of interest in all cases, except when 
the medium to be considered corresponds to the inner part of the inclusion. 
Fast calculations of the periodic Green's functions and of its gradient 
have been implemented as described in \cite{journal29}. A non-exhaustive list of alternative formulations for 
the periodic Green's function is given in \cite{handbook}. A more recent formulation
is described in \cite{volski2013}.

Hereafter, the {\bf Z} matrix is denoted by different symbols (from {\bf A} to {\bf E}), 
depending on the pairs of surfaces 
$S_1$ and $S_2$ considered for the interactions:
\begin{itemize}
\item{$\bf Z$ = $\bf A$ when $S_2$ is an interface and $S_1$ is also an interface.}
\item{$\bf Z$ = $\bf B$ when $S_2$ is boundary of inclusion and $S_1$ is interface.}
\item{$\bf Z$ = $\bf C$ when $S_2$ is interface and $S_1$ is boundary of inclusion.}
\item{$\bf Z$ = $\bf D$ when $S_2$ is boundary of inclusion and $S_1$ is boundary of inclusion,
in the host medium of the layer.}
\item{$\bf Z$ = $\bf E$ when $S_2$ is boundary of inclusion and $S_1$ is boundary of inclusion,
in the medium inside the object (non-periodic Green's function used).}
\end{itemize}
The two superscripts applied hereafter to those matrices will refer to the indices 
associated with surface $S_1$ and $S_2$, respectively (see Fig.~\ref{fig:layers}).
When $S_1$ and $S_2$ both correspond to interface $n$, the interaction can
take place through the host medium of layer $n$ or layer $n+1$; a
subscript will then indicate which layer should be considered.
The vectors of unknowns ${\bf x}_s^n$ and ${\bf x}_t^n$ will refer 
to the equivalent currents on interface $n$ and on the boundary of the inclusion
in layer $n$, respectively. 

Using the above notation, and excluding the first and last interfaces (i.e. $n \not= 0$ and $n \not=N-1$),
the continuity of tangential fields on interface $n$ is imposed through:
\begin{eqnarray}
{\bf A}^{n,n}_{n+1} {\bf x}_s^{n}  - {\bf A}^{n,n+1} {\bf x}_s^{n+1}  
+ {\bf B}^{n,n+1} {\bf x}_t^{n+1} &=& \nonumber \\
\hspace*{-4cm}
- {\bf A}^{n,n}_n {\bf x}_s^n  + {\bf A}^{n,n-1} {\bf x}_s^{n-1} + 
{\bf B}^{n,n} {\bf x}_t^{n}
\label{eq:cont_int}
\end{eqnarray}

The continuity of tangential fields on the boundary of the inclusion in layer $n$ is imposed through:
\begin{equation}
{\bf C}^{n,n} {\bf x}_s^{n} + {\bf C}^{n,n-1} {\bf x}_s^{n-1} + 
{\bf D}^{n,n} {\bf x}_t^{n} = - {\bf E}^{n,n} {\bf x}_t^{n}
\label{eq:cont_inc}
\end{equation}
The unknowns ${\bf x}_t$, referring to equivalent currents on the inclusions,
can be eliminated between (\ref{eq:cont_int}) written for interface $n$ 
and (\ref{eq:cont_inc}) written for the inclusions in layers $n$ and $n+1$.
A few straightforward algebraic transformations then yield:
\begin{equation}
{\bf A}_1^{n} {\bf x}_s^{n-1} + {\bf A}_2^{n} {\bf x}_s^{n} + 
{\bf A}_3^{n} {\bf x}_s^{n+1} = 0
\end{equation}
still for $n \not=0$ and $n \not=N-1$, and with the following matrices:
\begin{eqnarray}
{\bf A}_1^{n} &=& - {\bf A}^{n,n-1} - {\bf B}^{n,n}\, {\bf F}^{n} \label{eq:a1} \\
{\bf A}_2^n &=& \!\!\!{\bf A}^{n,n}_{n+1} \!+\! {\bf A}^{n,n}_n 
\!+\! {\bf B}^{n,n+1}\, {\bf F}^{n+1} \!-\! {\bf B}^{n,n}\, {\bf G}^{n} \label{eq:a2}\\
{\bf A}_3^{n} &=& - {\bf A}^{n,n+1} + {\bf B}^{n,n+1}\, {\bf G}^{n+1} \label{eq:a3}
\end{eqnarray}
and the following definitions
\begin{eqnarray}
{\bf F}^n &=& {\bf Y}^n\, {\bf C}^{n,n-1}\\
{\bf G}^n &=& {\bf Y}^n\, {\bf C}^{n,n}\\
{\bf Y}^n &=& - \left[ {\bf E}^{n,n} + {\bf D}^{n,n} \right]^{-1}
\end{eqnarray}
Section \ref{sec:numres} will show results for the case where inclusions are made
of core-shell nanoparticles, for which (\ref{eq:cont_inc}) needs to be generalized 
as explained in the appendix.

The special cases of $n=0$ and $n=N-1$ can be treated by setting
${\bf A}^{0,-1}=0$, ${\bf B}^{0,0}=0$, ${\bf A}^{N-1,N}=0$ and ${\bf  B}^{N-1,N}=0$,
in order to avoid contributions from non-existing boundaries. First, quite expectedly,
this leads to ${\bf A}_1^{0}=0$ and ${\bf A}_3^{N-1}=0$. Second, this also leads to
\begin{eqnarray}
\hspace*{-0.3cm}{\bf A}_2^0 &=& {\bf A}^{0,0}_1 + {\bf A}^{0,0}_0 
+ {\bf B}^{0,1}\, {\bf F}^{1}\\
\hspace*{-0.3cm}{\bf A}_2^{N-1} &=& \!\!\! {\bf A}^{N-1,N-1}_N \!\!+\! {\bf A}^{N-1,N-1}_{N-1} \!\! 
- \! {\bf B}^{N-1,N-1} {\bf G}^{N-1}
\end{eqnarray}
while ${\bf A}_3^{0}$ and ${\bf A}_1^{N-1}$ can be obtained from
(\ref{eq:a3}) and (\ref{eq:a1}), respectively.
The illumination from above only affects the condition to be satisfied at interface $N-1$,
for which an additional term ${\bf e}$ appears in the left hand side in (\ref{eq:cont_int}).
The vector ${\bf e}$ is the tested incident field:
\begin{equation}
\bf{e} = \left[
\begin{array}{c}
{\bf t}^e \\
{\bf t}^h
\end{array}
\right]
\end{equation}
with
\begin{equation}
{\bf t}^e(k) = \int_S\,\vec{F}_k\,.\,\vec{E}^{inc}\, dS
\end{equation}
where $\vec{E}^{inc}$ is the incident electric field and $\vec{F}_k$ is the $k$-th
basis function defined on interface $n=N-1$. A similar expression, involving
the incident magnetic field, is obtained for ${\bf t}^h(k)$.

Equations (\ref{eq:a1})-(\ref{eq:a3}) and the special cases given above
for layers $1$ and $N-1$ define a block tridiagonal system of equations:
\begin{equation}
\left[
\begin{array}{ccccc}
{\bf A}_2^0 & {\bf A}_3^0 & 0 & ...& 0\\
{\bf A}_1^1 & {\bf A}_2^1 & {\bf A}_3^1 & ... & 0\\
\vdots & & \ddots & & \vdots \\
0 & ... & 0 & {\bf A}_{1}^{N-1} & {\bf A}_{2}^{N-1}\\
\end{array}
\right]
\left[
\begin{array}{c}
{\bf x}_s^0 \\
{\bf x}_s^1 \\
\vdots \\
{\bf x}_s^{N-1}
\end{array}
\right]
=
\left[
\begin{array}{c}
0 \\
0 \\
\vdots \\
-{\bf e}
\end{array}
\right]
\end{equation}

The sparse matrix results from the use of the surface equivalence principle, in that equivalent currents eliminate
the direct interactions with inclusions and interfaces that are located beyond the immediately neighboring interfaces.
Also, the elimination of the unknowns on the inclusions represent a huge time saving since complex inclusions may require a very 
large number of basis functions (and hence of unknowns) for an accurate representation of the fields on their boundaries. 
The latter can, however, always be obtained a posteriori from the equivalent currents 
on the interfaces, by isolating ${\bf x}_t^n$ in (\ref{eq:cont_inc}). 
The unknowns describing the equivalent electric and magnetic currents on a given interface can be limited to a small number, typically ranging from 100 to 400, i.e. the number of entries in ${\bf x}_s$ typically ranges from  $M$=200 to 800. 
Besides, if $N$ is the number of interfaces, specialized solvers for block tridiagonal matrices can be used \cite{numrec}; their complexity is simply proportional to $N$. Hence, the complexity of the solution process is of the order of $M^3\, N$. Considering, for instance, a problem with 8 interfaces discretized with 256 basis functions (i.e. 512 unknowns per interface), independently from the complexity of the inclusions, 
will take the same time as the inversion of a square matrix of dimension 1024, i.e. a fraction of a second on a present-day desktop computer. 

\begin{figure*}[t]
\begin{center}
\includegraphics[scale=0.5]{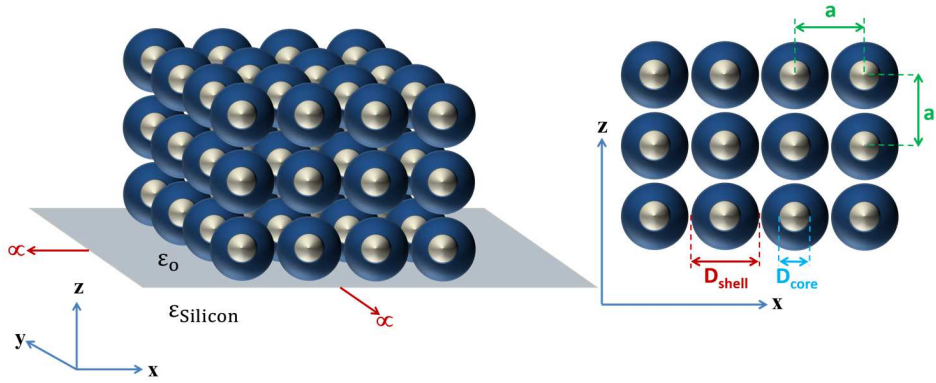}
\end {center}
\caption{The geometric details of the array configuration are given for the 3-layer, infinite doubly periodic array of core-shell nanoparticles.}
\label{fig:Array_Geometry}
\end{figure*} 

\section{Numerical Results}
\label{sec:numres}
\indent{In this section, we validate the proposed numerical approach for a multilayered, infinite doubly periodic array of 
core-shell nanoparticles above a silicon substrate. This configuration has been studied in \cite{journal34} in an attempt 
to realize low-index metamaterials using a newly developed self-assembling technology. 
The array is excited by an $x$-polarized plane wave propagating in the $-z$ direction (see Fig.~\ref{fig:Array_Geometry}). 
The reflectivity of the array is obtained in the frequency range which corresponds to free-space wavelength 
 $\lambda_o$ ranging from $300$ nm to $700$ nm. 
The diameters of the silver core and silica shell are $26$ nm and $82$ nm, respectively. 
The periods of the array in $x$ and $y$ directions are both equal to $84$ nm. 
The layers of arrays of core-shell nanoparticles in the $z$ direction are separated by $2$ nm.
Fig.~\ref{fig:Array_Geometry} illustrates the geometrical details of the array configuration. 
In this configuration, the lowest layer (i.e. layer 0) is taken as a semi-infinite space of silicon,
the upper layer (i.e. layer $N$) is made of air, and the host media of the $N-1$ intermediate layers are air.
The refractive indices of silicon and fused silica are taken from \cite{list1} and \cite{book2}, respectively. 
The refractive index of silver is based on Palik's measurements with a correction term \cite{palik}. 
Those refractive indices are converted into complex relative permittivities assuming that the complex
relative permeability is 1. Two different background media are considered and the core-shell
inclusions have a certain level of complexity in both geometry and material composition. 
Hence, this structure forms an interesting example to validate the numerical method proposed here.

\begin{figure}[htb]
\begin{center}
\includegraphics[scale=0.28]{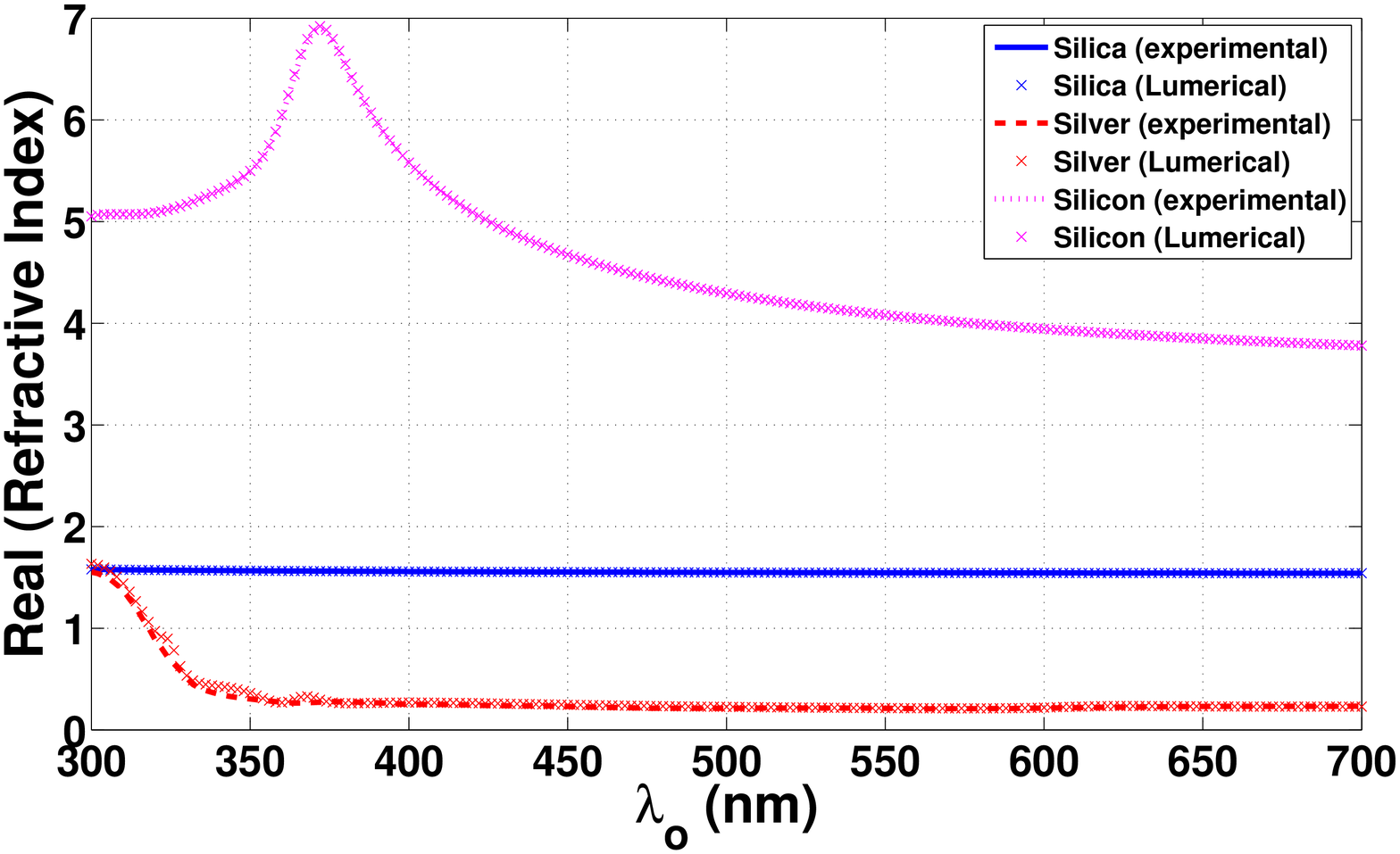}
\caption{Real part of refractive index of three materials used}
\label{fig:realn}
\end{center}
\end{figure}

\begin{figure}[htb]
\begin{center}
\includegraphics[scale=0.28]{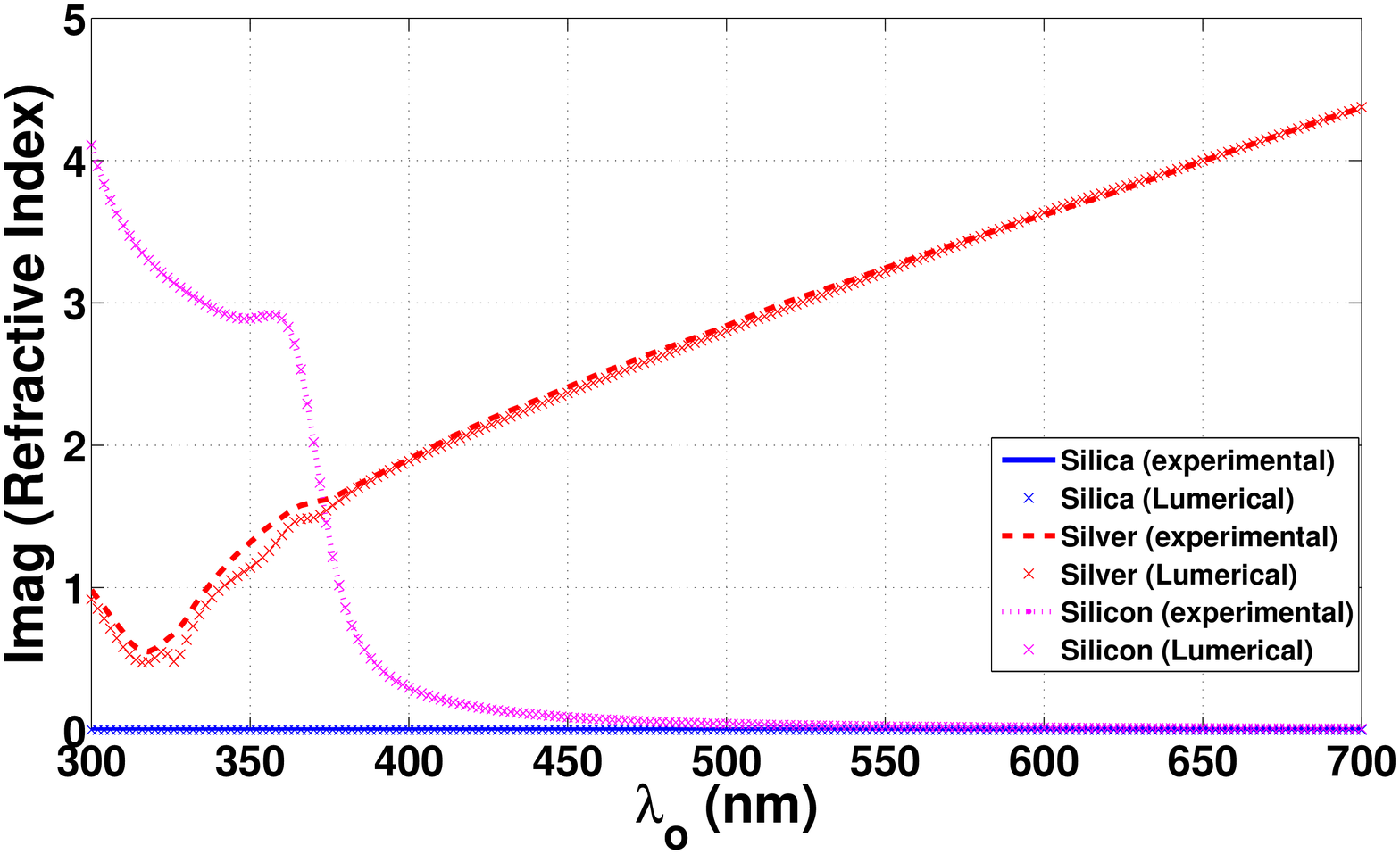}
\caption{Imaginary part of refractive index of three materials used}
\label{fig:imagn}
\end{center}
\end{figure}

The surfaces of the silver core and of the silica shell are both discretized into 486  RWG \cite{rwg82} basis functions. 
The interface between silicon substrate and free space is discretized into 392 rooftop basis functions. 
The interface between the layers of particles in free space is discretized into 72 rooftop basis functions. 
The Lumerical FDTD solutions \cite{lum} software has been employed
to produce the reference results. 
Lumerical has been used with the high mesh accuracy parameter and a minimum mesh step of 1 nm. 
A proper comparison between MoM and FDTD results requires
a closer inspection of material parameters. 
Lumerical employs multi-coefficient models that multiply a set of basis functions to better fit dispersion profiles.
For high accuracy frequency-domain outputs, this may not be sufficient to fit the
experimental refractivity index data to the desired accuracy (e.g. with an error below 0.05).
Hence, in this work, in order to obtain a better validation of accuracy of the MoM results,
the highest number of coefficients has been used for the refractivity model used in Lumerical, 
and the corresponding model has been exploited in the MoM simulations.
Figures \ref{fig:realn} and \ref{fig:imagn} show the real and imaginary parts of the 
refractive indices of silver, silicon, and silica, as from experiments \cite{palik}, \cite{list1}, \cite{book2},
and as resulting from the Lumerical model referred to above.}

\begin{figure}[htb]
\begin{center}
\includegraphics[scale=0.28]{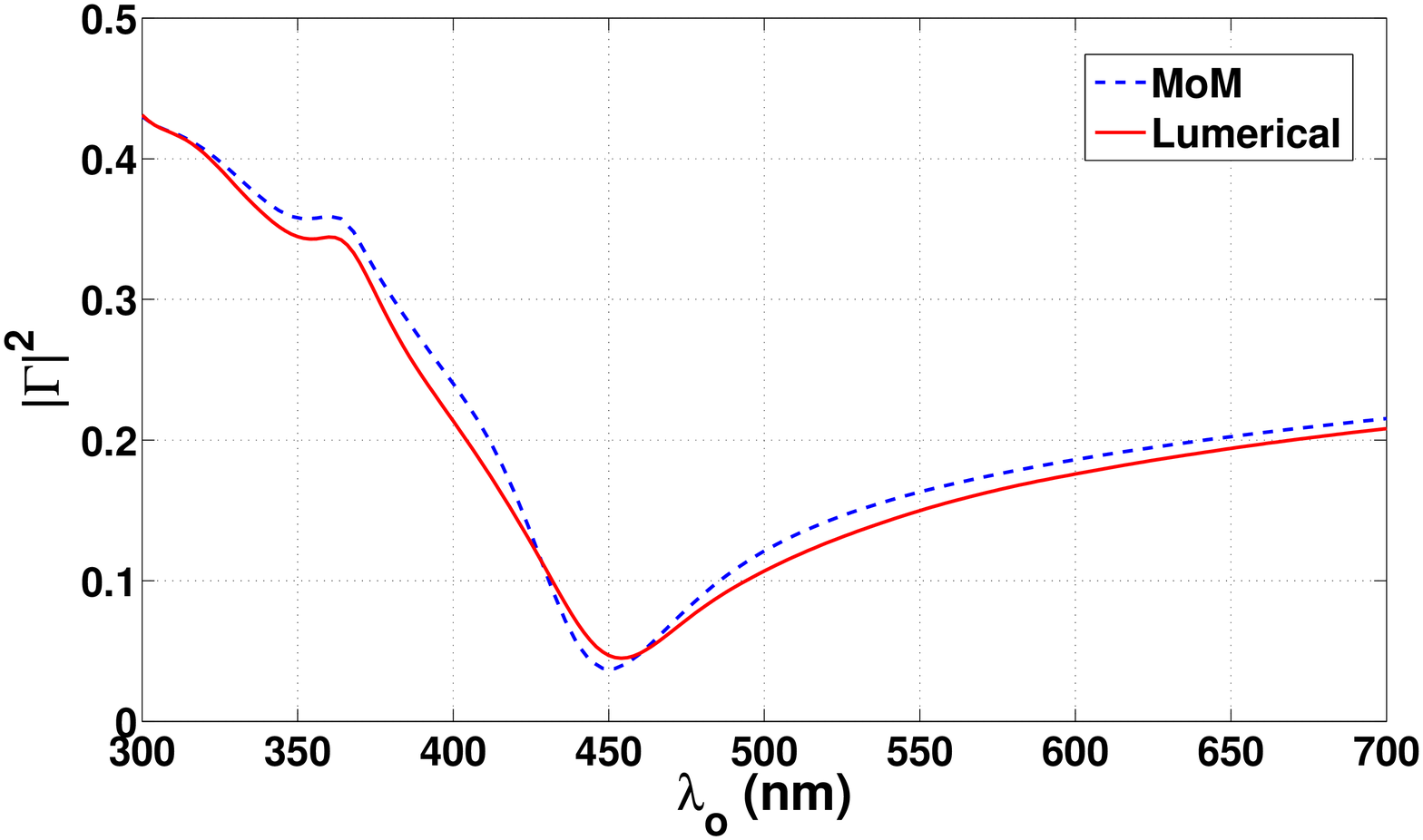}
Fig.~{\ref{fig:results}.a:  Reflectivity for one layer.}
\end{center}
\end{figure}
\begin{figure}[htb]
\begin{center}
\includegraphics[scale=0.28]{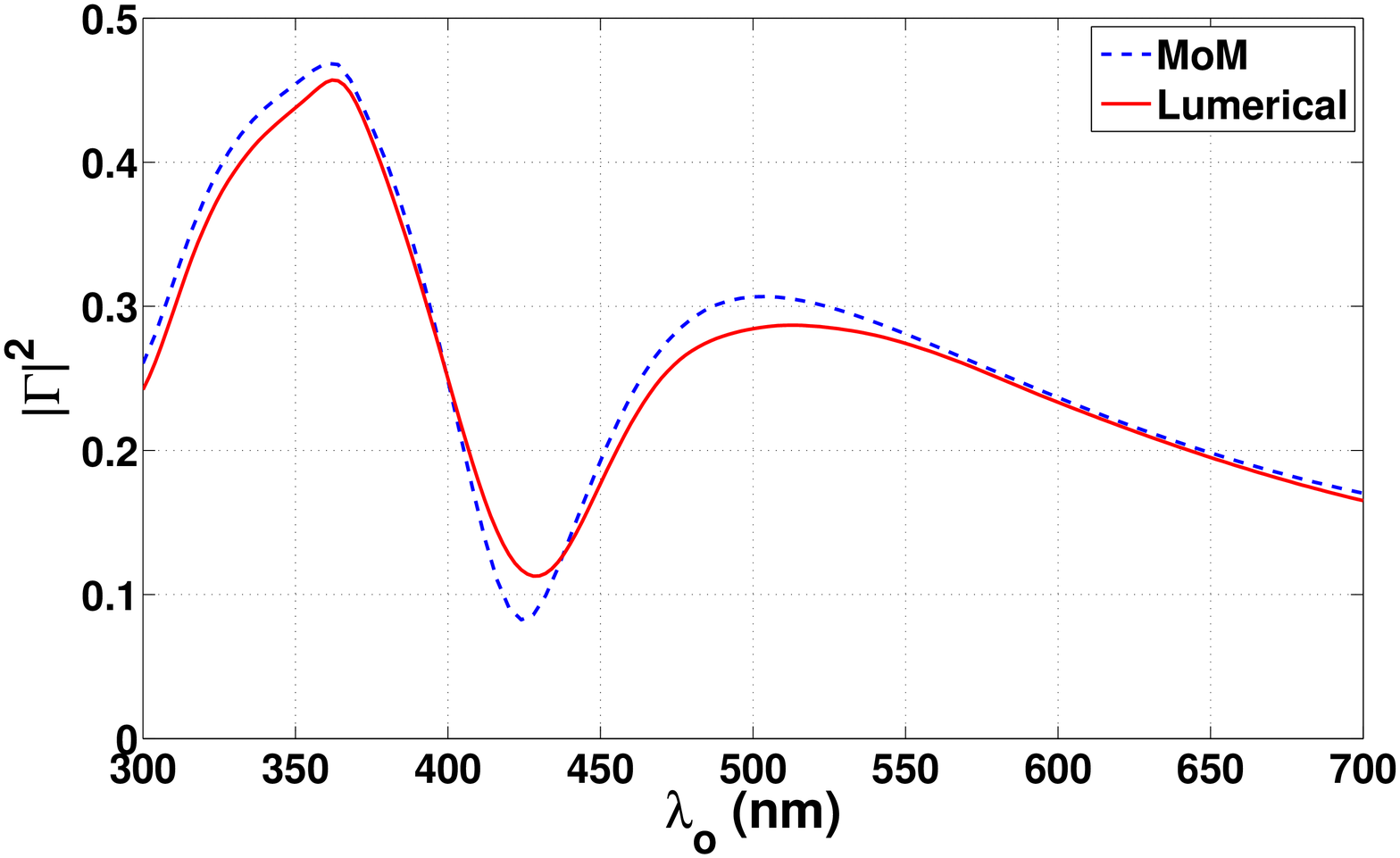}
Fig.~{\ref{fig:results}.b:  Reflectivity for two layers.}
\end{center}
\end{figure}
\begin{figure}[htb]
\begin{center}
\includegraphics[scale=0.28]{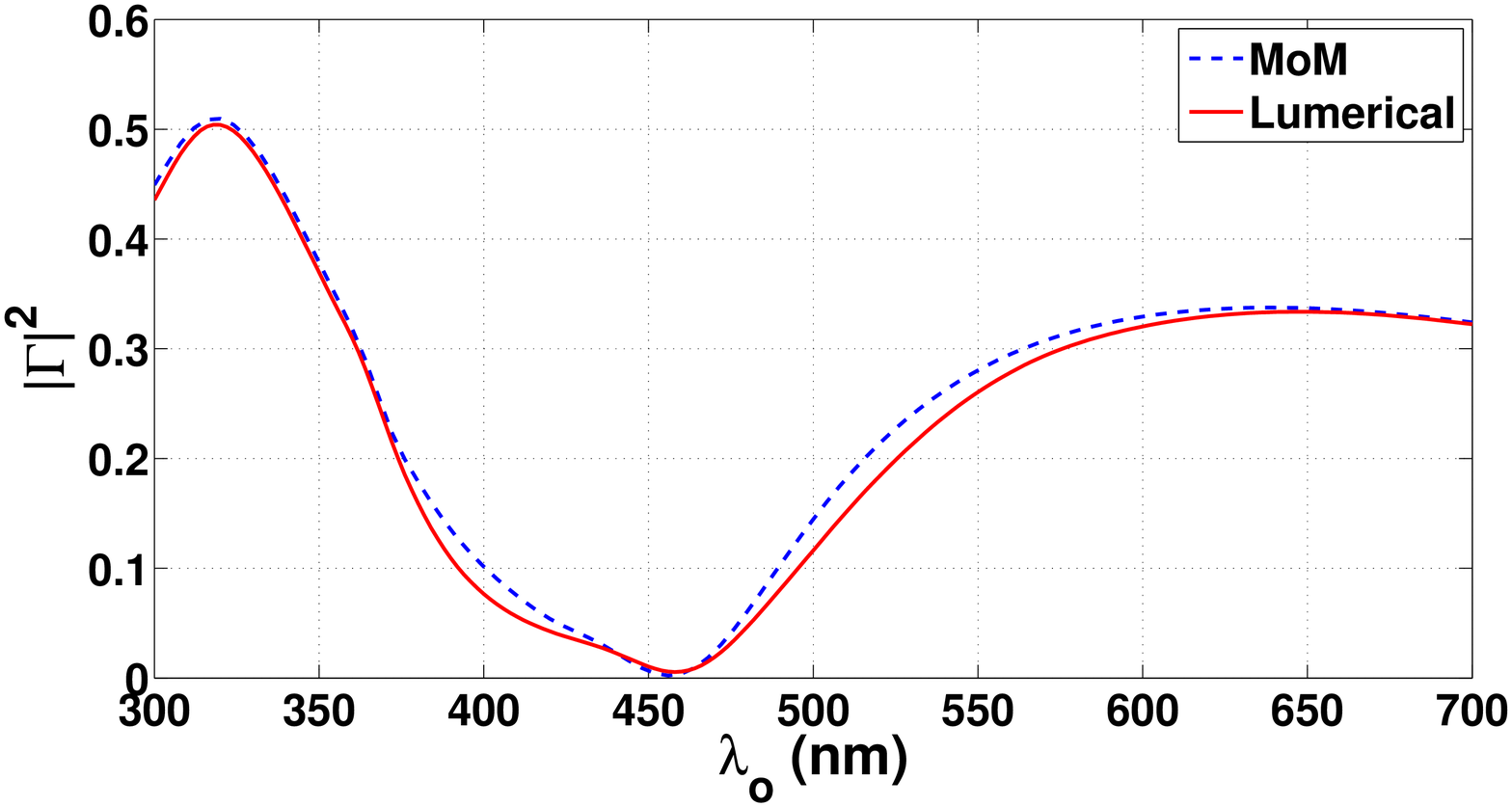}
Fig.~{\ref{fig:results}.c:  Reflectivity for three layers.}
\end{center}
\end{figure}
\begin{figure}[htb]
\begin{center}
\includegraphics[scale=0.28]{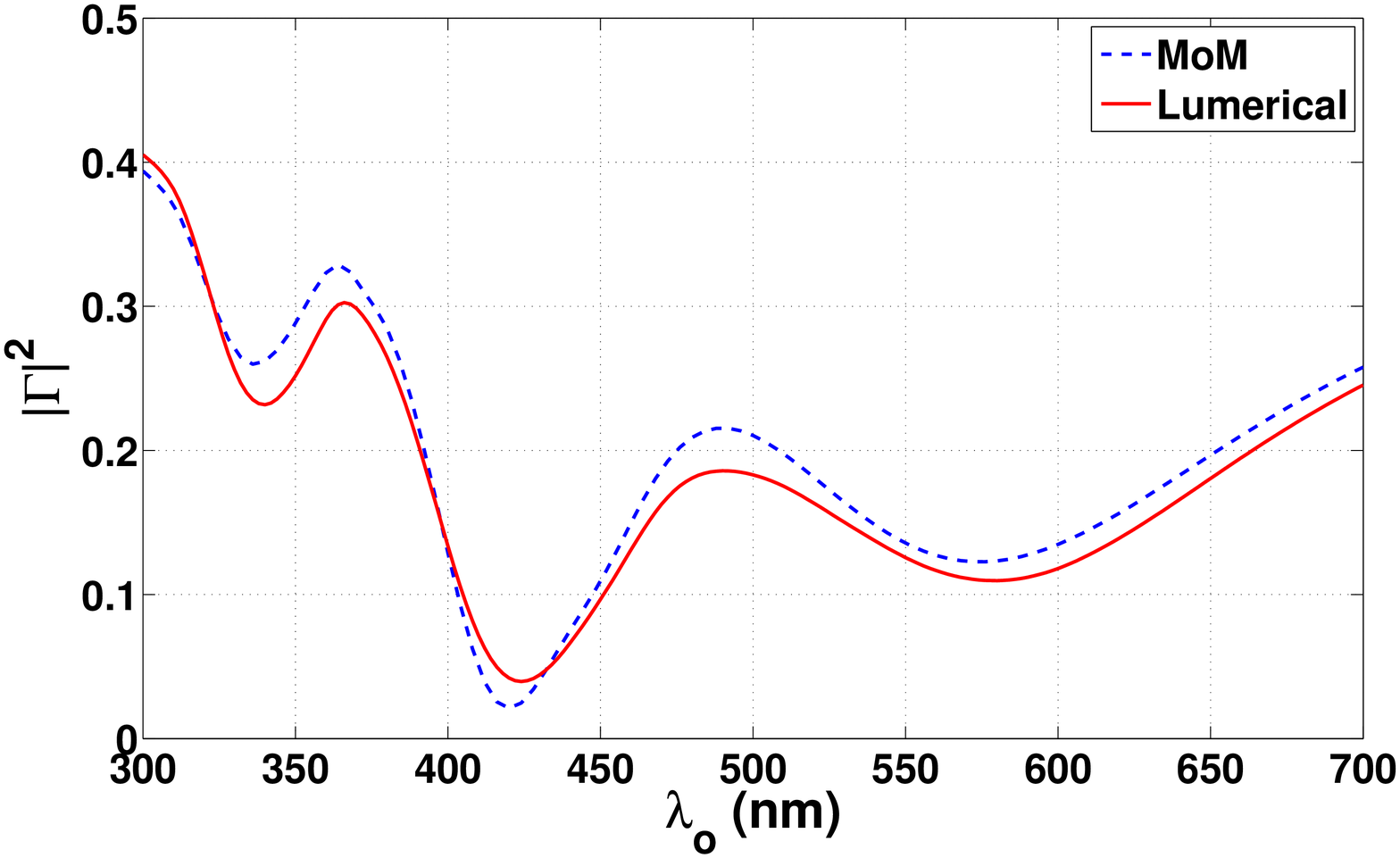}
Fig.~{\ref{fig:results}.d:  Reflectivity for four layers.}
\end{center}
\end{figure}
\begin{figure}[htb]
\begin{center}
\includegraphics[scale=0.28]{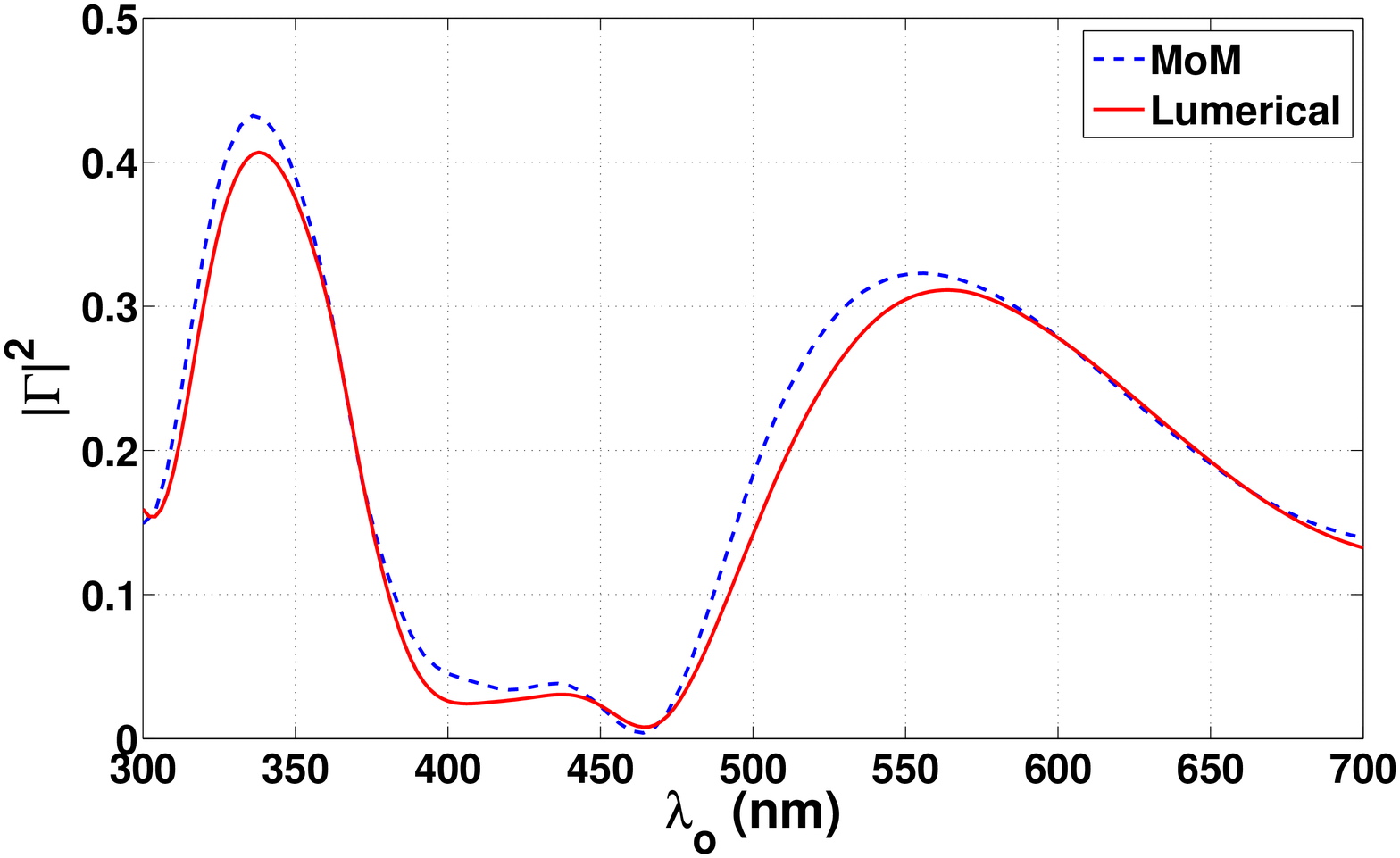}
Fig.~{\ref{fig:results}.e:  Reflectivity for five layers.}
\end{center}
\end{figure}
\begin{figure}[htb]
\begin{center}
\includegraphics[scale=0.28]{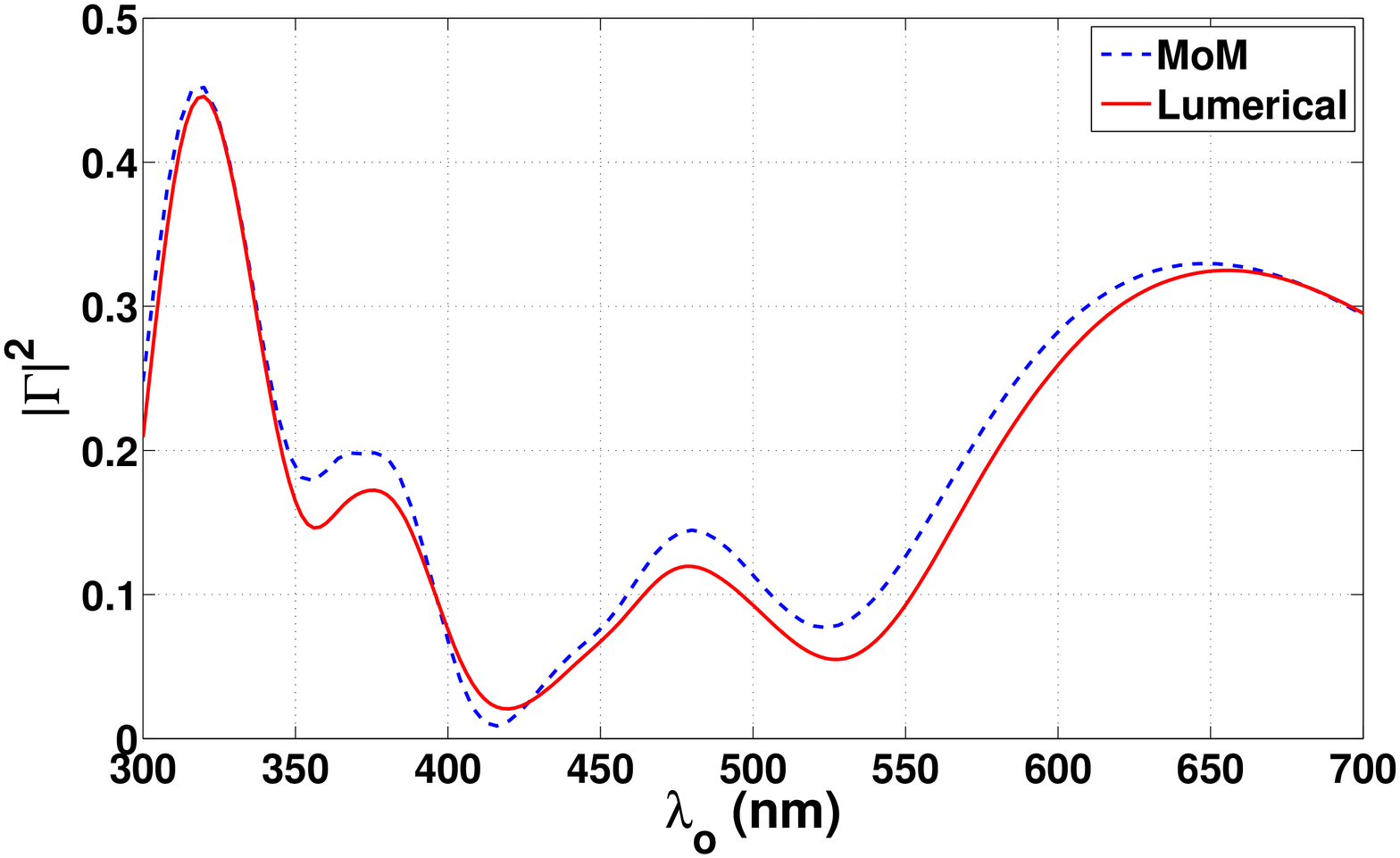}
Fig.~{\ref{fig:results}.f:  Reflectivity for six layers.}
\end{center}
\end{figure}
\begin{figure}[htb]
\begin{center}
\includegraphics[scale=0.28]{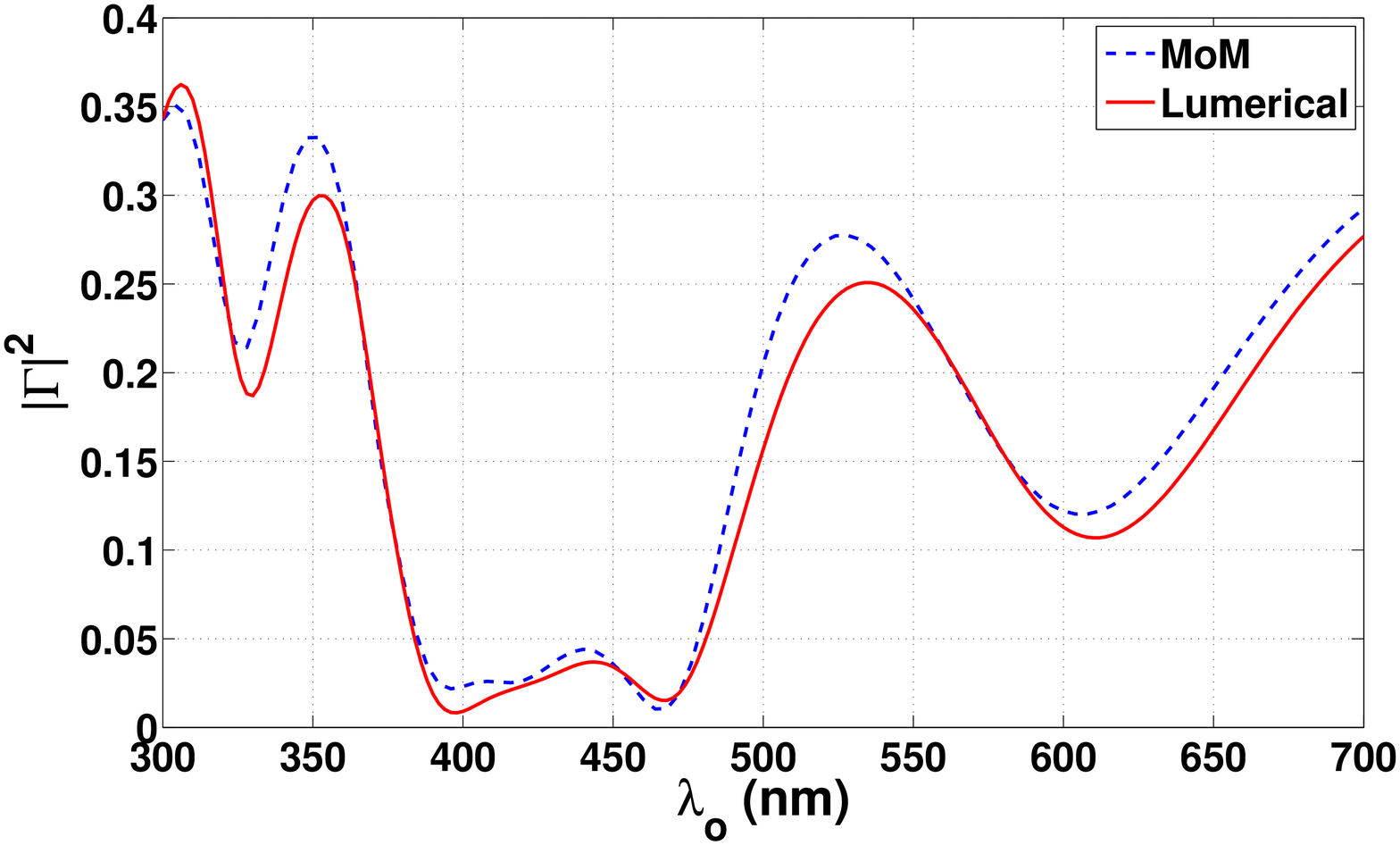}
Fig.~{\ref{fig:results}.g:  Reflectivity for seven layers.}
\caption{Reflectivity for an increasing number of layers
of silver-silica core-shell nano-particles above a silicon substrate. This MoM
approach in dashed lines and Lumerical solution in solid lines.}
\label{fig:results}
\end{center}
\end{figure}

Reflectivity results, i.e. the square-magnitude of the reflection coefficient, 
over the [300-700] nm wavelength range are shown in plots $a$ to $g$ of Fig.~\ref{fig:results}.
A very good agreement between reflectivities obtained using the proposed
approach and Lumerical is observed at all frequencies
and for all numbers of layers. For a number of layers larger than three
and for high reflectivity values (for $|\Gamma|^2$ larger than about 0.15), 
the reflectivity predicted with our approach tends to be slightly higher than 
that predicted with Lumerical: a shift of the order of 0.02 is observed,
although one does not observe a systematic shift, since there is virtually no
shift in some sub-bands. Our investigations with both MoM and FDTD
simulations did not allow us to determine which of both both sets of results 
is most accurate. 

Hereafter, computation times will be given for simulations carried out
on a PC with Intel Core processor i7−4770K CPU with 3.5 GHz clock rate.
On that computer, the Lumerical software uses 8 processors, 
while the MoM software uses only one, because the MoM code has
not been parallelized yet.
Hence, in order to compare, computation times will be converted to equivalent
times for a single-processor operation, i.e. they will be multiplied by 8 for
the FDTD computations and left as is for the MoM calculations.
One should note that this multiplication by 8 assumes a good parallelization scaling
of the Lumerical software. The MoM approach can be strongly accelerated by exploiting 
interpolation of MoM matrices (i.e. the different types of ${\bf Z}$ matrices referred to
in Section \ref{sec:form}) versus frequency \cite{newman}. This leads to very important time savings,
since -even for seven layers- the computation time is by far dominated
by the filling of the MoM matrices. Here, over the frequency band
of interest, the different ${\bf Z}$ matrices are computed explicitly at only
11 of the 101 frequencies and second-order interpolation is used, without significant loss 
of accuracy at other frequencies (the maximum deviation for one layer in terms of
reflectivity $|\Gamma|^2$ is near 0.004, observed near the resonant peak of the silicon 
refractive  index at 365 nm). The computation times are given in three columns in Table \ref{tab}:
$(i)$ MoM matrices obtained explicitly at 11 frequencies, $(ii)$ all other computations 
involved in the MoM solution and $(iii)$ FDTD calculations in terms of 
single-processor equivalent (8 times the solution time observed
with parallel computation on 8 cores).
\begin{table}
  \caption{Computation times in seconds: MoM matrix computations,
all other MoM computations and FDTD computations assuming only
one processor}
  \centering
  \begin{tabular}{|c|c|c|c|}
    \hline
    \# & MoM matrices & MoM other &  FDTD \\
    \hline
    1 layer & 1060  & 66 & 17760\\
    2 layers  & 1285 & 70 & 19200\\
    3 layers  & 1285 & 79 & 19200\\
    5 layers  & 1285 & 90 & 21220\\
    7 layers & 1285 &  98 & 26880\\
    \hline
 \end{tabular}
 \label{tab}
\end{table}

Taking into account that FDTD caculations have been carried out
on eight cores, it can be seen that MoM calculations are faster
by slightly more than one order of magnitude (more precisely by
factors of 15.8, 14.2 and 19.4 for 1, 2 and 7 layers, respectively).
Moreover, the MoM calculation time is by far dominated by
the matrix filling time, which is never larger than the one
needed for two layers; also, the total computation time
for seven layers exceeds that for two layers by only 28 seconds,
while considering the 101 frequencies at which reflectivity is computed. 
Finally, the MoM matrix filling computations for different frequencies could 
also be trivially distributed over several processors. The relatively large
time involved in matrix filling underscores the relevance of
further research in the acceleration of calculations of periodic Green's 
functions and MoM interactions between elementary basis functions.
Regarding the latter, our implementation could be optimized in
several respects.

\section{Conclusions}

A new formulation has been presented for the simulation of scattering
by multi-layered structures with periodic penetrable inclusions contained in each layer,
and assuming common periods in the plane of the layers. 
The method requires a minimal set of calculation routines, corresponding to electric or magnetic-type 
interactions between basis and testing functions, using a 2-D periodic Green's function
in unbounded homogeneous space. The principal characteristic of the method 
is the introduction of interstitial equivalent currents at the interfaces, which
allows the decoupling of non-adjacent layers. In this way, the use
of Green's functions associated with multilayered media is avoided,
the unknowns associated with the complex inclusions can be eliminated
from the final system of equations; and the complexity
grows only linearly with the number of layers.

Validating numerical results have been provided for the case of scattering by 
stacked layers of core-shell nano-particles placed above a silicon substrate.
A very good agreement has been observed, as compared with results obtained
with a commerical software based on FDTD, while computation times are shorter
by a factor of the order of 16 for the MoM calculations.
Given this efficient treatment of the multi-layered aspect of the structures,
it becomes clear that further research should focus on the acceleration of 
MoM matrix-filling techniques for the relatively fundamental problem of periodic
structures in unbounded media. Further research also concerns
implementations with non-planar interfaces, which may, for instance, be
considered in the case of hexagonal arrangements, as well as
the possibility of refining the mesh at interfaces, specificially in regions
extremely close to the inclusions.

\section*{Acknowledgement}
This work has been funded by the EU METACHEM project and by
the Belgian BESTCOM project and by the Belgian Interuniversity 
Attraction Poles Programme 7/23 BESTCOM.

\section*{Appendix}
In the case where a secondary inclusion is embedded into a primary inclusion,
as considered in Section \ref{sec:numres}, where the silica inclusion contains
a silver core, expression (\ref{eq:cont_inc}) needs to be generalized. Let us denote
by ${\bf x}_i^{n}$ the equivalent current on the outer boundary of
the secondary inclusion, and by $S_t$ and $S_i$ the surfaces that enclose
the primary and secondary inclusions, respectively. Then we can write:
\begin{eqnarray}
{\bf C}^{n,n} {\bf x}_s^{n} + {\bf C}^{n,n-1} {\bf x}_s^{n-1} +
{\bf D} {\bf x}_t^{n} \!\!&=&\!\! - {\bf E} {\bf x}_t^{n} + {\bf M}_{t,i} {\bf x}_i^{n} \label{eq:cont_inc1}\\
{\bf M}_{i,t} {\bf x}_t^{n} +  {\bf M}_{i,i} {\bf x}_i^{n} \!\!&=&\!\! - {\bf N} {\bf x}_i^{n} 
\label{eq:cont_inc2}
\end{eqnarray}
where ${\bf M}_{a,b}$ is a $\bf Z$-type matrix standing for interaction, through
the medium between $S_t$ and $S_i$, between basis functions located
on $S_b$ and testing functions on surface $S_a$, where indices $a$ and $b$ can
both correspond to either $t$ or $i$ indices. Matrix $\bf N$ has the same definition
as matrix $\bf E$, except that the surface is that of the secondary inclusion and that
the medium considered is the one inside that inclusion. The other matrices have been
defined in Section \ref{sec:form}. To alleviate notations, superscripts $n,n$ have been
removed from the matrices that were wearing them. Then, the current on the
secondary inclusion can be eliminated between equations (\ref{eq:cont_inc1}) 
and (\ref{eq:cont_inc2}), to obtain an equation that has exactly the same form as (\ref{eq:cont_inc}),
assuming a new definition for matrix $\bf E$.

\ifCLASSOPTIONcaptionsoff
  \newpage
\fi

%








\end{document}